\documentclass[conference]{IEEEtran}
\IEEEoverridecommandlockouts

\usepackage[utf8]{inputenc}

\usepackage{cite}
\usepackage{amsmath,amssymb,amsfonts}
\usepackage{algorithmic}
\usepackage{graphicx}
\usepackage{textcomp}
\usepackage{xcolor}
\def\BibTeX{{\rm B\kern-.05em{\sc i\kern-.025em b}\kern-.08em
    T\kern-.1667em\lower.7ex\hbox{E}\kern-.125emX}}
\begin{document}

\title{Desarrollo de experiencias para la enseñanza y difusión de la Óptica con impresión 3D\\
\thanks{
Todos los autores pertenecen también al Centro de Investigaciones Ópticas (CONICET – UNLP – CIC), Gonnet, Argentina.}
}


\author{
\IEEEauthorblockN{Roberto Peyton}
\IEEEauthorblockA{
\textit{Depto. de Ciencia y Tecnología}\\
\textit{Universidad Nacional de Quilmes}\\
Bernal, Argentina\\
robertop@ciop.unlp.edu.ar}
\and
\IEEEauthorblockN{Damián Presti}
\IEEEauthorblockA{
\textit{Depto. de Ciencia y Tecnología}\\
\textit{Universidad Nacional de Quilmes}\\
Bernal, Argentina\\
damianp@ciop.unlp.edu.ar}
\and
\IEEEauthorblockN{Jeffry H. Martínez Valdiviezo}
\IEEEauthorblockA{
\textit{Depto. de Ciencia y Tecnología}\\
\textit{Universidad Nacional de Quilmes}\\
Bernal, Argentina\\
jeffryh@ciop.unlp.edu.ar}
\and
\IEEEauthorblockN{Fabián Videla}
\IEEEauthorblockA{
\textit{Facultad de Ingeniería}\\
\textit{Universidad Nacional de La Plata}\\
La Plata, Argentina\\
fabianv@ciop.unlp.edu.ar}
\and
\IEEEauthorblockN{Gustavo Adrian Torchia}
\IEEEauthorblockA{
\textit{Depto. de Ciencia y Tecnología}\\
\textit{Universidad Nacional de Quilmes}\\
Bernal, Argentina\\
gustavot@ciop.unlp.edu.ar}
}

\IEEEoverridecommandlockouts
\IEEEpubid{\makebox[\columnwidth]{978-1-7281-5957-7/20/\$31.00~\copyright2020 IEEE \hfill} \hspace{\columnsep}\makebox[\columnwidth]{ }}

\maketitle

\IEEEpubidadjcol

\begin{abstract}
In this work, we present three experiences developed with 3D printing technology for teaching in Optics. These activities were planned, designed, and implemented to go deeper into knowledge about light polarization, reflection and refraction phenomena, and colour perception. The planning methodology is also detailed, in particular, the design and description of the activity. Experimental guides, 3D models and assembly manuals are available in a free and freely accessible web space. In summary, we demonstrate that the 3D printing can be used as a technological, inclusive and pedagogical resource for teaching. This methodology can be extended and adapted to other disciplines, as well as applied locally in educational institutions.\\
\end{abstract}


\renewcommand\abstractname{Resumen}
\begin{abstract}
En este trabajo se presentan tres experiencias desarrolladas con la tecnología de impresión 3D para la enseñanza y difusión de la Óptica. En particular, se planificaron, diseñaron e implementaron actividades que abordan los conceptos de polarización de la luz, los fenómenos de reflexión y refracción en una interfaz y la percepción del color. En cada caso se detalla la metodología utilizada en la planificación, profundizando sobre el diseño y la descripción de la experiencia. La documentación necesaria para llevar a cabo las prácticas,  guías experimentales, modelos 3D y manuales de armado, se encuentran disponibles en un espacio web de acceso libre y gratuito.  El uso de la impresión 3D es un recurso tecnológico apropiado para desarrollar nuevas   aptitudes en el proceso de aprendizaje y  criterios pedagógicos. Por otra parte la realización de experiencias de laboratorio desarrolla actitudes inclusivas, desde un enfoque participativo. Esta metodología puede extenderse y adecuarse a otras disciplinas, como así también aplicarse concretamente en las instituciones locales de enseñanza.\\
\end{abstract}


\section{Introducción}
La impresión 3D es uno de los avances tecnológicos más importantes de los últimos tiempos, ha revolucionado la manera de fabricar objetos reales a partir de su concepción virtual, siendo esos prototipos faciles de modificar y/o mejorar. Estas características han despertado interés  y propiciado su adopcion en  todo el mundo  \cite{b1,b2,b3}. El éxito de ésta tecnología radica en su versatilidad, flexibilidad, y principalmente en su bajo costo, transformándose así en una disciplina asequible para la mayoría de la población. La impresión 3D consiste en la creación de un objeto tridimensional, modelado en un programa de diseño CAD, a través de un proceso aditivo de sucesivas capas. En este proceso se pueden utilizar diversos tipos de materiales: plástico, metal, tejidos biológicos, cerámica, hormigón, etc. \cite{b2}. Ésta tecnología ha introducido avances y mejoras en distintos campos de aplicación, por ejemplo en el sector industrial para diseños de prototipos, en la medicina para la creación de prótesis, en la industria aeroespacial, en aplicaciones científicas, entre otros \cite{b3}; y la educación no ha sido ajena a esta tendencia. En los últimos años, se han realizando una amplia variedad de investigaciones del uso de la impresión 3D como herramienta pedagógica, en todos los niveles del sistema educativo y en diversas áreas del conocimiento \cite{b4}. Por otra parte, el avance de las nuevas tecnologías ha generado un cambio de paradigma en las metodologías de enseñanza, a punto tal, que hoy en día no se concibe una práctica educativa en la cual no se utilice algún tipo de medio tecnológico, sean aulas virtuales, medios audiovisuales, redes sociales, u otros \cite{b5}; en efecto, es menester promover la utilización de la impresión 3D como recurso tecnológico en la educación.

En particular, es de gran interés desarrollar a nivel regional instrumentos para la enseñanza y difusión de la Óptica con impresión 3D, principalmente, porque la mayoría de los elementos utilizados en estos kits educativos deben ser importados, limitando la planificación y encareciendo la implementación de prácticas experimentales \cite{b6,b7}. Este tipo de instrumental es clave para una adecuada comprensión de los distintos fenómenos estudiados por esta disciplina, como así también para generar un espacio participativo entre educandos y conocimiento. Cada una de las actividades experimentales, además, pueden ser adecuadas a las necesidades particulares de las instituciones locales de enseñanza. Por ejemplo, uno de los desarrollos en la región latinoamericana relacionados con la enseñanza de la Óptica, es el llevado a cabo por la Universidad Privada Boliviana de Cochabamba. Consiste en un espectrómetro óptico fabricado por los mismos alumnos en impresión 3D. Utiliza la cámara de un teléfono móvil para la captura de datos y una aplicación para ejecutar ajustes y mediciones de forma sencilla. Toda la información es de acceso público y gratuito desde la web \cite{b8,b9}. En consecuencia, este dispositivo es un claro ejemplo del potencial que ofrece esta tecnología como herramienta pedagógica y de inclusión en la enseñanza de la ciencia.

En este trabajo desarrollamos una serie de dispositivos experimentales utilizando la tecnología de impresión 3D con aplicación en la enseñanza y difusión de la Óptica, pensadas para distintos niveles del sistema educativo (secundario-universitario). Para ello diseñamos y fabricamos una variedad de piezas y accesorios comúnmente utilizados en los laboratorios de esta disciplina para el montaje de las experiencias. Además, planificamos cada una de las experiencias con el objetivo de profundizar en un tema específico del conocimiento, en cada caso, proponemos una guía teórico-práctica para llevar a cabo las actividades. Finalmente, utilizamos un espacio web de acceso libre y gratuito para distribuir todo el material necesario, es decir, guías experimentales, modelos 3D de cada una de las piezas y manuales de armado de las experiencias.

\section{Materiales y métodos}

En primer instancia se conformó un grupo colaborativo compuesto por docentes, investigadores y estudiantes para planificar los experimentos a desarrollar. En segunda instancia se diseñaron e implementaron las experiencias. En particular, para el modelado de las piezas y accesorios se utilizaron las herramientas de diseño asistido por computadora FreeCAD y Blender. La trayectorias se generaron con el programa CreatWare. Las impresiones de las piezas se realizaron con una impresora 3D Createbot DE, que tiene un recorrido de 400x300x300 mm. Se utilizó ácido poli-láctico (PLA) como material, este polímero, además de ser biodegradable, debido a su bajo punto de fusión puede utilizarse en todas las impresoras existentes en el mercado \cite{b10}. Por otra lado, se diseño una plaqueta electrónica con el programa KiCad, la cual se fabricó con el ampliamente conocido método de estampado de m\'{a}scara impresa con  plancha. Adicionalmente, se emplearon variados accesorios para completar las experiencias, en su mayoría de fácil acceso en el mercado local, como por ejemplo punteros láser, tornillos, impresiones en papel, piezas de acrílico, componentes de electrónica, fuentes de alimentación para tiras led, entre otros. Todos estos elementos se especifican detalladamente en los manuales de armado de las experiencias.

Finalmente, se organizaron una serie de jornadas, en donde docentes, investigadores y estudiantes participaron en conjunto de la planificación y confección de las actividades, las guías experimentales y los manuales de armado de las experiencias didácticas. Todo este material es de acceso público y gratuito desde la web, para ello se utilizó un dominio provisto por la Universidad Nacional de Quilmes \cite{b11}.

\section{Propuesta}

Se eligieron algunos contenidos de interés en Óptica, profundizando en cada caso en los fundamentos teóricos. Posteriormente, teniendo en cuenta los sujetos de aprendizaje que participarán de las actividades, se proyectaron tres experimentos que pudieron ser implementados, en su mayoría, con piezas realizadas en impresión 3D. A continuación se ordenan de menor a mayor, según la complejidad de elaboración, los temas elegidos:
\begin{itemize}
\item Polarización.
\item Reflexión y refracción.
\item Percepción del color.
\end{itemize}

El objetivo de la experiencia de polarización es investigar la naturaleza electromagnética de la luz.  El fenómeno de polarización resulta apropiado a ese fin, pese al relativo  desconocimiento que el público tiene sobre él, es un ejemplo aplicado y cotidiano; en consecuencia es de esperar que su ``descubrimiento''  aliente la  curiosidad del alumno y además incremente sus saberes prácticos. La polarización es utilizada, por ejemplo, en pantallas de tecnología LED, gafas de sol, protectores solares en automóviles, filtros para cámaras, entre otros. En particular, se emplean dos contenidos conceptuales específicos, la ley de Malus y la fotoelasticidad. Se experimenta con distintos materiales y fuentes de luz (lámparas incandescentes, LED, láseres y pantallas de dispositivos). La ley de Malus describe las variaciones experimentadas por el campo electromagnético al atravesar dos filtros polarizadores lineales. La fotoelasticidad en cambio, es una técnica óptica que permite visualizar esfuerzos y deformaciones mecánicas en materiales transparentes.

La experiencia sobre reflexión y refracción tiene como propósito la enseñanza de la óptica geométrica (teoría de rayos), como así también el alcance de esta teoría para explicar ciertos aspectos de la naturaleza de la luz. En particular se utiliza el fenómeno de refracción y reflexión en una interfaz, conceptos que se describen matemáticamente por la ley de Snell. Una vez abordado y aplicado el concepto, se presentan algunos de los casos en donde el trazado de rayos no alcanza para explicar el comportamiento físico, como por ejemplo el ángulo de polarización, reflectancia y transmitancia, contraponiendo de esta forma la óptica geométrica y la óptica física (electromagnética). Por lo tanto, con esta metodología asimismo podemos mostrar y explicar cómo diferentes paradigmas y teorías científicas entraron en conflicto a lo largo de la historia.

Finalmente, la experiencia sobre percepción del color tiene como objetivo darle las herramientas a los estudiantes para que den los primeros pasos en la comprensión del color, pero desde un abordaje científico. Desde esta perspectiva, el color como tal, no existe, solo es una percepción o la interpretación que el cerebro hace de las señales nerviosas que le envían los ojos. Para profundizar en esta teoría, se necesita esencialmente una fuente de luz que pueda modificar sus colores de salida fabricado con impresión 3D. Concretamente, los alumnos recrearán colores basándose en la síntesis aditiva, del mismo modo que lo hacen todas las pantallas de tecnología LED. Podríamos pensar que, desde el punto de vista didáctico, el experimento imita la típica experiencia de la niñez de mezclar pinturas de diferentes colores.

\section{Resultados}

\begin{figure}[htbp]
\centerline{\includegraphics{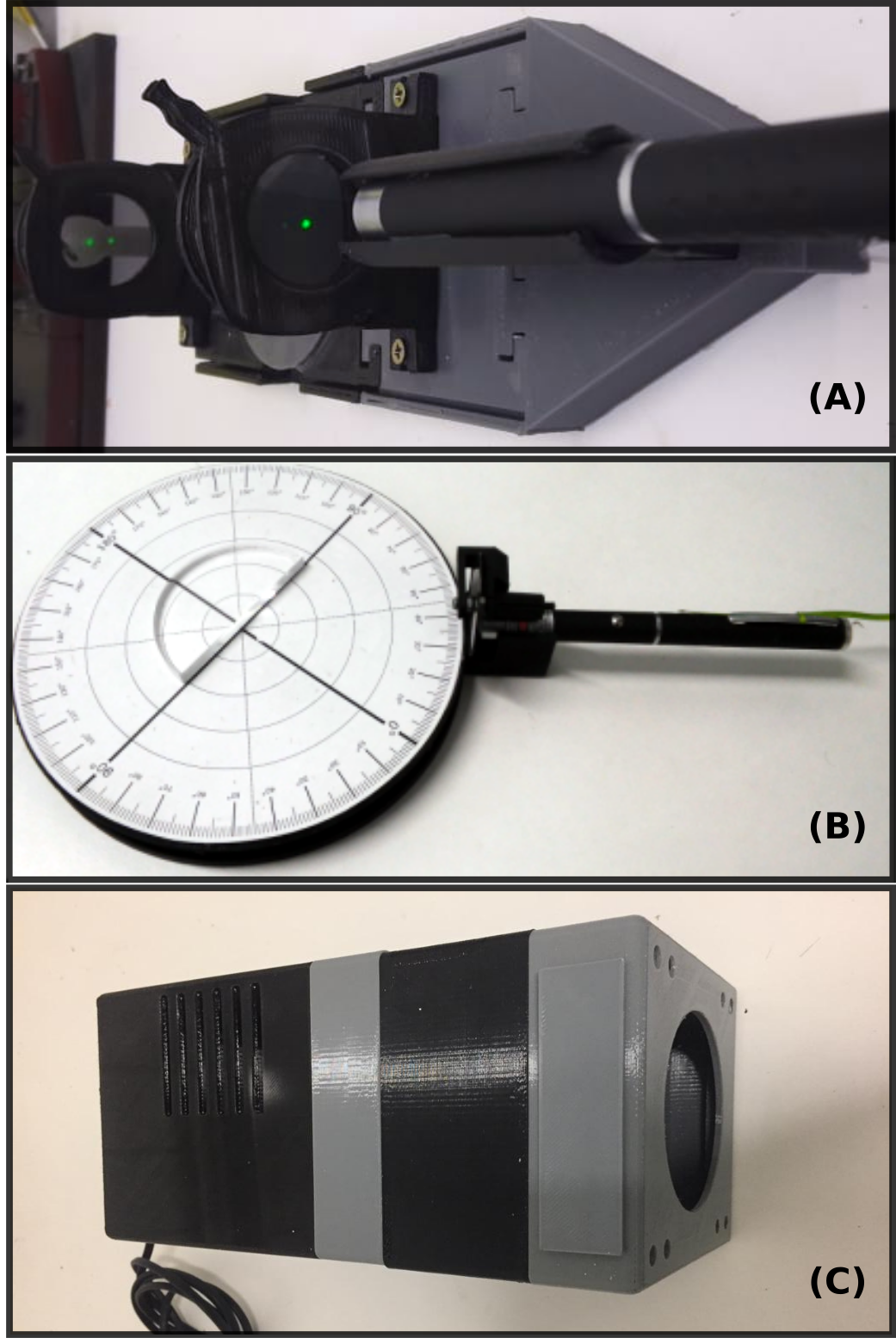}}
\caption{Fotografías de los tres equipos desarrollados con la tecnología de impresión 3D. Se indican para cada uno los objetivos experimentales: (A) polarización; (B) reflexión y refracción; (C) percepción del color.}
\label{fig0}
\end{figure}

En esta sección mostramos los tres dispositivos implementados. Además se detallan en cada caso las fortalezas y debilidades de los sujetos de aprendizaje, como así también el conocimiento abordado y actividades propuestas. También se describe el diseño, fabricación y armado de las experiencias con la tecnología de impresión 3D. Por último, presentamos la estructura utilizada en las guías experimentales y las metodologías de evaluación propuestas. Antes de profundizar en las particularidades de cada diseño, en la Fig.~\ref{fig0} se presentan los tres equipos desarrollados.

\subsection{Polarización}

\begin{figure}[htbp]
\centerline{\includegraphics{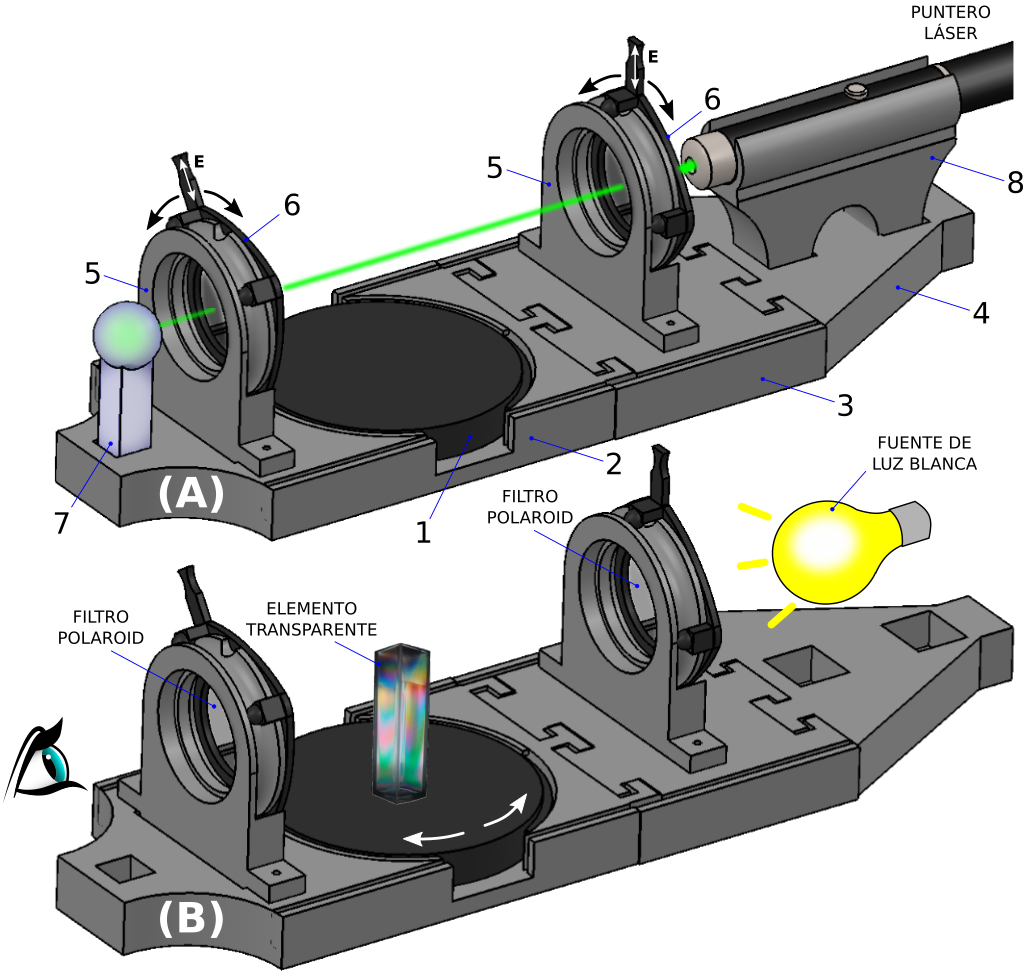}}
\caption{Esquema de experiencia sobre polarización. Las piezas numeradas corresponden a las fabricadas con impresión 3D, las demás corresponden a los elementos comerciales. En la figura se muestra las actividades sobre (A) ley de Malus y (B) fotoelasticidad.}
\label{fig1}
\end{figure}

En la Fig.~\ref{fig1} se muestran los dos esquemas experimentales planificados, ambos emplean un dispositivo cuyo diseño está compuesto por una plataforma y dos soportes giratorios con polarizadores tipo polaroid, todas las piezas numeradas corresponden a las fabricadas con impresión 3D. La experiencia podrá ser armada por los mismos alumnos al momento de iniciar la actividad, desarrollando de este modo un espíritu lúdico en el proceso de aprendizaje. Se hace notar que todas las partes del experimento son acoplables, y en particular, la plataforma tiene una geometría de ``rompecabezas". 

El dispositivo para explorar la ley de Malus se exhibe en la Fig.~\ref{fig1} (A), en esta experiencia se utiliza un haz láser que atraviesa dos polarizadores e incide en una pieza dispersante. La ley se pone de manifiesto al cambiar la posición angular relativa de los polarizadores, se observará un máximo de transmisión cuando los polarizadores estén paralelos y un mínimo cuando estén perpendiculares. Por otro lado, en la Fig.~\ref{fig1} (B) se esboza la actividad sobre fotoelasticidad, en este caso, se emplea una pieza de plástico transparente apoyada sobre la base giratoria y una fuente de luz blanca en un extremo. La actividad consiste en alinear los polarizadores y observar desde el otro extremo la pieza. Aquí se apreciará una gama de colores (irisado) que revela la distribución de tensiones mecánicas. Luego, al rotar la pieza, se observará un cambio de la distribución inicial.

\subsection{Reflexión y refracción}

\begin{figure}[htbp]
\centerline{\includegraphics{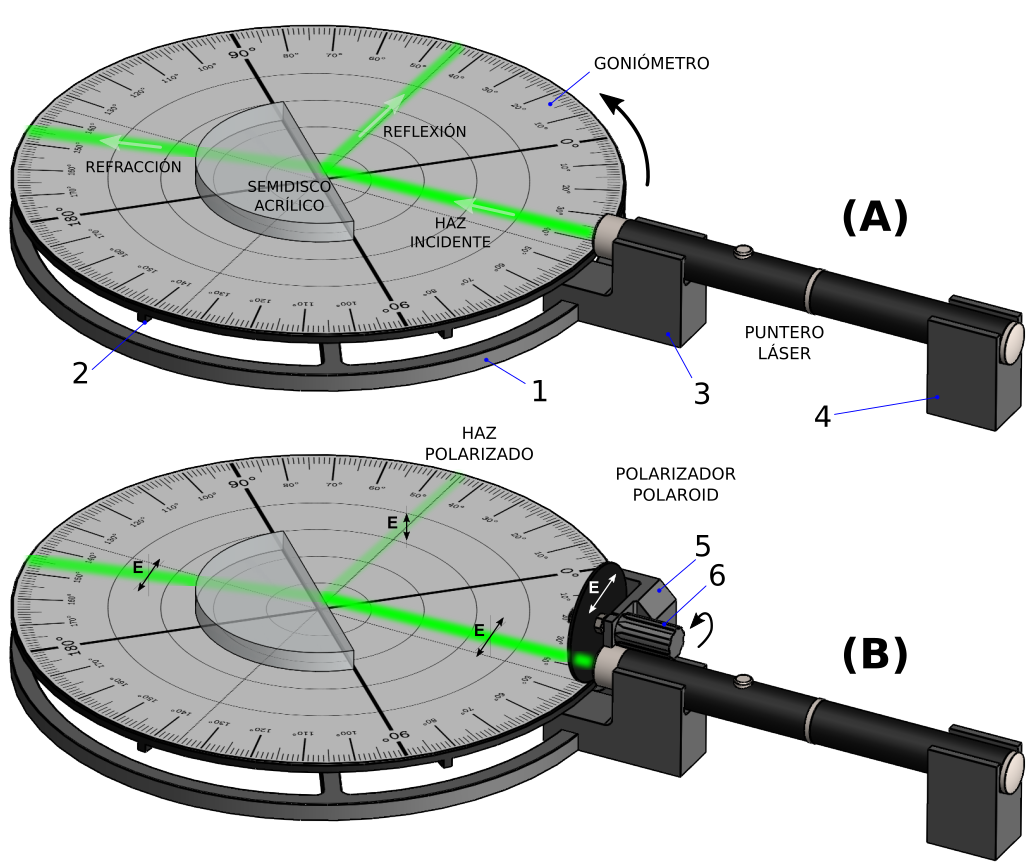}}
\caption{Esquema de experiencia sobre reflexión y refracción. Las piezas numeradas corresponden a las fabricadas con impresión 3D, las demás corresponden a los elementos comerciales. En la figura se muestra las actividades sobe (A) óptica geométrica y (B) óptica física.}
\label{fig2}
\end{figure}

El diseño consiste en una base giratoria donde se apoya un semi-disco de acrílico pulido, y un puntero láser fijado a un lado de la base. Los ángulos de giro pueden ser medidos utilizando un goniómetro impreso en papel. En la Fig.~\ref{fig2} se expone un esquema de las experiencias. Para las tareas referidas a la óptica física, se agrega un soporte giratorio con un polarizador que permite fijar la orientación del campo electromagnético del haz. La experiencia, al igual que en el caso anterior, puede ser armada por los propios alumnos al momento de iniciar la actividad. Adicionalmente, se requiere realizar un proceso de alineación entre el semi-disco y el láser, desarrollando de esta forma un procedimiento típico de los laboratorios de óptica.

En la primer actividad se experimenta con la ley de Snell, tal como se muestra en el esquema de la Fig.~\ref{fig2} (A). En especial, se verifican las relaciones entre los ángulos reflejados y refractados. Para esto se utiliza la reflexión externa, cuando el índice de refracción del medio incidente es menor que el índice del medio transmisor, y el caso contrario, es decir, reflexión interna. Con estos ensayos y utilizando la ley de Snell, se propone calcular y verificar el índice de refracción del acrílico y el ángulo de reflexión total interna. Por otra parte, en la segunda actividad se experimenta con los efectos electromagnéticos. Por este motivo se procede a utilizar luz polarizada. La actividad consiste en calcular y verificar el ángulo de polarización o de Brewster, tal como se muestra en el esquema de la Fig.~\ref{fig2} (B). Finalmente, se analiza cualitativamente la reflectancia y transmitancia para las diferentes polarizaciones.

\subsection{Percepción del color}

\begin{figure}[htbp]
\centerline{\includegraphics{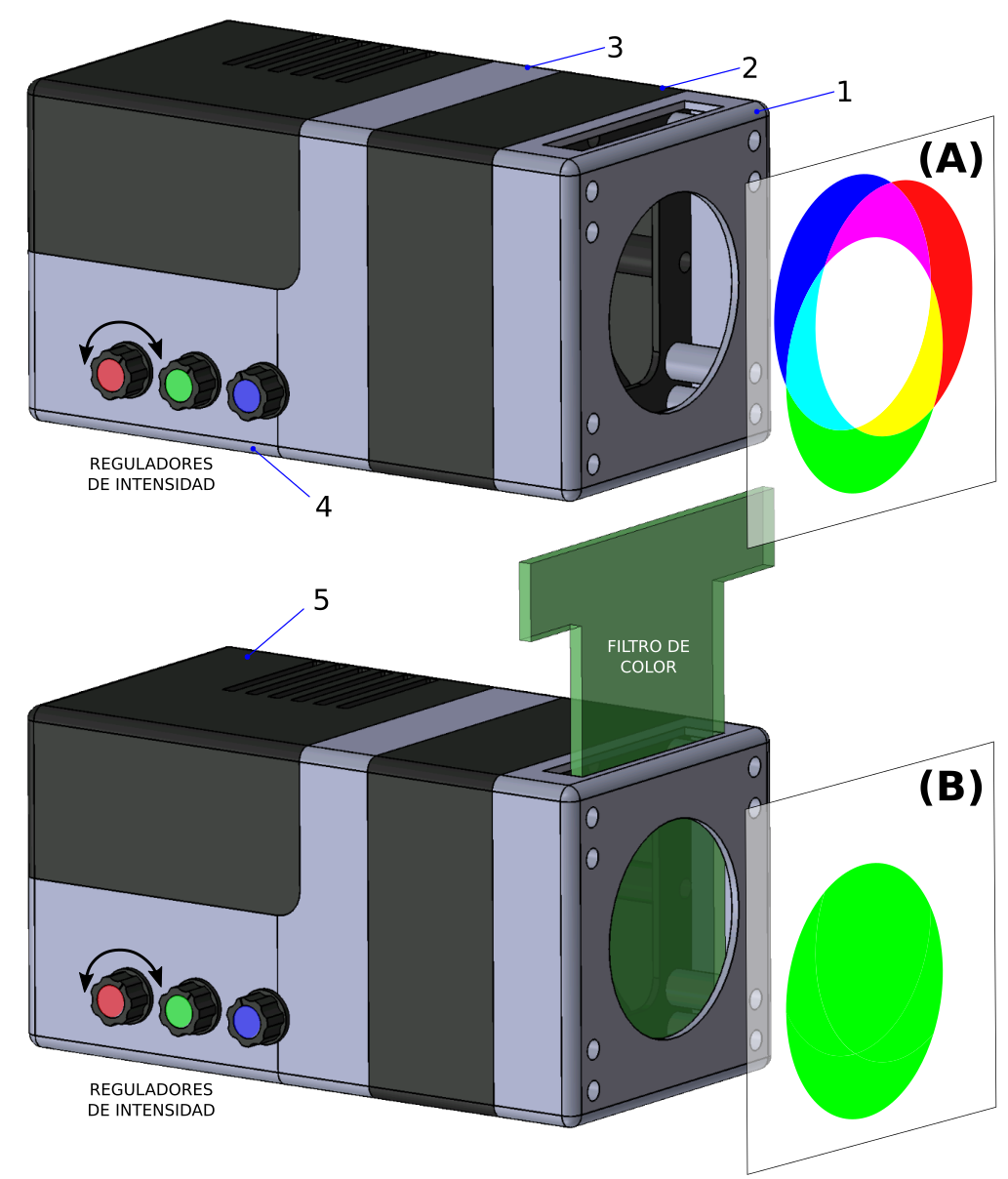}}
\caption{Esquema de experiencia sobre la percepción del color. Las piezas numeradas corresponden a las fabricadas con impresión 3D, las demás corresponden a los elementos comerciales. En la figura se muestra las actividades sobe (A) síntesis aditiva y (B) síntesis sustractiva.}
\label{fig3}
\end{figure}

Como se puede ver en la Fig.~\ref{fig3}, el diseño propuesto es fundamentalmente un proyector basado en el modelo RGB. Este equipo permite reproducir colores por síntesis aditiva utilizando tres leds de colores separados (rojo, azul y verde), cada led puede variar su intensidad mediante reguladores de corriente. De este modo, utilizando una pantalla donde se proyecta la luz, se puede visualizar el esquema de colores de la síntesis aditiva ya mencionada. Además, se deja a disposición un receptáculo que permite interponer diversos elementos de interés, como por ejemplo filtros de colores, difusores, etc.

La actividad se inicializa con la exploración, es decir, se propone que cada estudiante investigue y asocie los diferentes componentes que generan un color. Para ello, se giran los reguladores de intensidad contemplando todas las combinaciones posibles; concluyendo con la formación del color blanco, tal como se dibuja en la Fig.~\ref{fig3} (A). Asimismo, se procede a utilizar diferentes elementos en la proyección, con el objeto de explicar las teorías de colorimetría. En la Fig.~\ref{fig3} (B) se expone un esquema modelo, en este ejemplo se utiliza un filtro de color verde. Por último, se plantea como actividad la generación de colores a partir de su composición RGB, verificando y ajustando las proporciones con un dispositivo móvil. Para esto existen múltiples aplicaciones gratuitas, por ejemplo Color Grab \cite{b12}.

\subsection{Documentación de las experiencias}

Se ha elaborado una serie de plantillas que sirven como base para replicar las experiencias. Las guías experimentales y los manuales de armado pueden ser utilizados tanto por alumnos como docentes. Cabe destacar que la documentación fue estructurada siguiendo los métodos científicos. En este sentido, los alumnos además de experimentar y hacerse de nuevos conocimientos, serán partícipes de una típica práctica de laboratorio. Por otra parte, las guías para actividades experimentales, junto con los manuales de armado de cada kit y los planos de las piezas a ser fabricados mediante impresión 3D, han sido subidos a una página web, dentro del repositorio de la Universidad Nacional de Quilmes \cite{b11}.

Las guías de actividades experimentales contienen la siguiente información:
\begin{itemize}
\item Título: refleja la actividad a desarrollar así como las magnitudes físicas que se determinarán en el experimento.
\item Objetivos: muestra claramente cuáles son los propósitos  y alcances de la experiencia.
\item Marco Teórico: se describen los fundamentos físicos y modelos presentes en el experimento, incluyendo la bibliografía correspondiente.
\item Materiales: se especifican los materiales, instrumentos y piezas fabricadas en impresión 3D.
\item Procedimiento experimental: se explican los pasos procedimentales y métodos de ajuste.
\item Resultados: curvas, gráficos y/o tablas a utilizar para el registro y análisis de datos.
\item Discusión y conclusiones: se propone un análisis crítico sobre los resultados y actividades de investigación para que los participantes formulen sus propias conclusiones.
\end{itemize}

\begin{figure}[htbp]
\centerline{\includegraphics{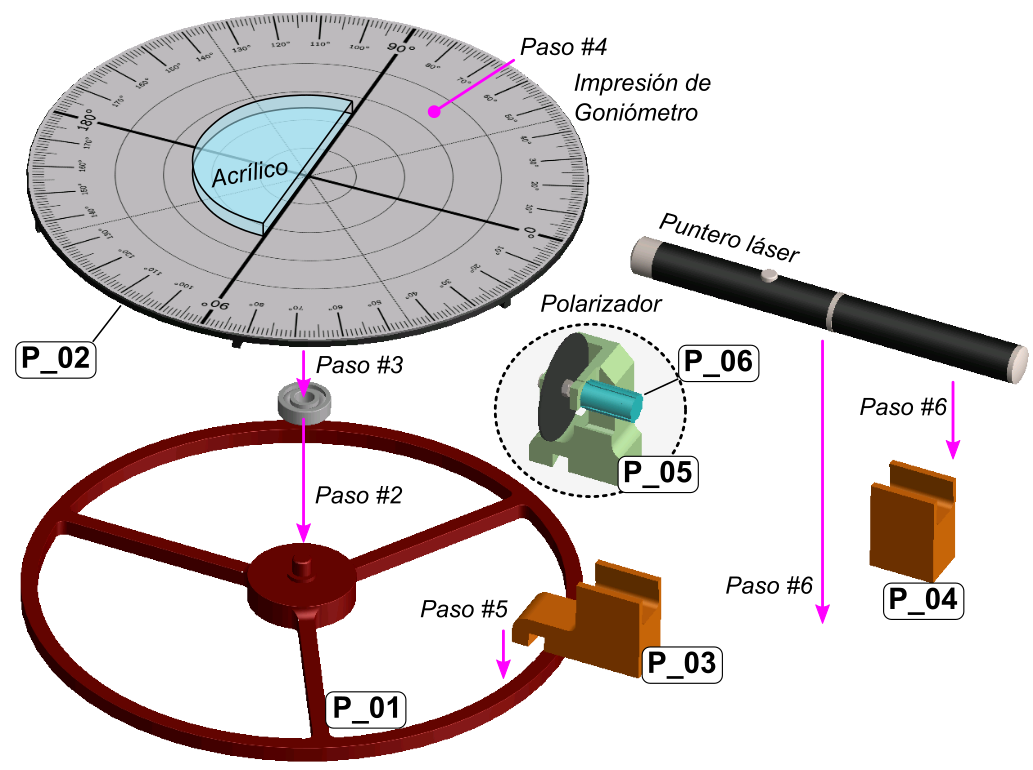}}
\caption{Esquema del despiece utilizado en el manual de armado de la experiencia sobre reflexión y refracción.}
\label{fig4}
\end{figure}

En los manuales de armado se especifican los materiales, componentes y piezas a fabricar en impresión 3D. Asimismo, se aclara en detalle cada uno de los pasos a seguir para el correcto armado, acompañado en cada caso de imágenes que ayudan a entender el proceso. En la Fig.~\ref{fig4} se muestra, a modo de ejemplo, el esquema del despiece utilizado en el manual de armado de la experiencia sobre reflexión y refracción. 

\section{Conclusiones}

En este trabajo presentamos la producción de tres experiencias útiles para la enseñanza y difusión de la Óptica realizadas con impresión 3D. Las actividades fueron planificadas conjuntamente por docentes, investigadores y estudiantes; considerando que estas puedan ser implementadas por los propios alumnos en las instituciones educativas. En este sentido, todos los materiales utilizados en los diseños se caracterizan por ser componentes comerciales de fácil acceso en el mercado local. Elaboramos además la documentación necesaria para llevar a cabo las prácticas, organizada así: guías experimentales, modelos 3D y los manuales de armado. Todo este material se encuentra disponible en un espacio web de acceso libre y gratuito \cite{b11}.
En resumen, mostramos tres ejemplos del uso de la impresión 3D como recurso tecnológico, de inclusión y pedagógico para la enseñanza, en el que los participantes recorren activamente,  guiados por el material elaborado,  varios procesos de aprendizaje, desde la elaboración de la actividad hasta la formación del conocimiento. Este enfoque puede extenderse y adecuarse a otras disciplinas, contemplando las necesidades particulares de las instituciones locales de enseñanza. 

\section*{Agradecimientos}

Este trabajo ha sido posible gracias al financiamiento de la Universidad Nacional de Quilmes a través del proyecto PPROF-901-2018, la Agencia Nacional de Promoción Científica y Tecnológica (ANPCyT) a través de los proyectos PICT-2016-4086 y PICT-2017-0017. Roberto Peyton, Damián Presti y Gustavo A. Torchia pertenecen al Consejo Nacional de Investigaciones Científicas y Técnicas (CONICET); Jeffry H. Martínez Valdivieso agradece a la Agencia Nacional de Promoción Científica y Tecnológica (ANPCyT) por la financiación la beca financiada; y Fabian Videla pertenece a la Comisión de Investigaciones Científicas de la Provincia de Buenos Aires (CIC-BA).


\begin{thebibliography}{00}
\bibitem{b1} R. Jiang, R. Kleer, and F. T. Piller, “Predicting the future of additive manufacturing: A Delphi study on economic and societal implications of 3D printing for 2030,” Technological Forecasting and Social Change, vol. 117, pp. 84–97, Apr. 2017.
\bibitem{b2} J.-Y. Lee, J. An, and C. K. Chua, “Fundamentals and applications of 3D printing for novel materials,” Applied Materials Today, vol. 7, pp. 120–133, Jun. 2017.
\bibitem{b3} T. D. Ngo, A. Kashani, G. Imbalzano, K. T. Q. Nguyen, and D. Hui, “Additive manufacturing (3D printing): A review of materials, methods, applications and challenges,” Composites Part B: Engineering, vol. 143, pp. 172–196, Jun. 2018.
\bibitem{b4} S. Ford and T. Minshall, “Invited review article: Where and how 3D printing is used in teaching and education,” Additive Manufacturing, vol. 25, pp. 131–150, Jan. 2019.
\bibitem{b5} R. M. Hernandez, “Impacto de las TIC en la educación: Retos y Perspectivas,” Propósitos y Representaciones, vol. 5, no. 1, p. 325, Apr. 2017.
\bibitem{b6} Newport Corporation. 'Laboratory Educational Kits', 2020. [Online]. Available: https://www.newport.com/c/laboratory-educational-kits. [Accessed: 15-November-2020].
\bibitem{b7} Thorlabs, Inc. 'Physics Lab Experiments for College', 2020. [Online]. Available: https://www.thorlabs.com/navigation.cfm?guide\_id=2310. [Accessed: 15-November-2020].
\bibitem{b8} O. Ormaechea, A. Villazón, and R. Escalera, “A spectrometer based on smartphones and a low-cost kit for transmittance and absorbance measurements in real-time,” Optica Pura y Aplicada, vol. 50, no. 3, pp. 239–249, Sep. 2017.
\bibitem{b9} Universidad Privada Boliviana. 'Espectrómetro para Teléfonos Inteligentes', 2017. [Online]. Available: http://www.upb.edu/es/contenido/espectrometro-para-telefonos-inteligentes. [Accessed: 15-November-2020].
\bibitem{b10} M. A. Elsawy, K.-H. Kim, J.-W. Park, and A. Deep, “Hydrolytic degradation of polylactic acid (PLA) and its composites,” Renewable and Sustainable Energy Reviews, vol. 79, pp. 1346–1352, Nov. 2017.
\bibitem{b11} Universidad Nacional de Quilmes. 'Kits educativos de óptica', 2020. [Online]. Available: http://fotonica.web.unq.edu.ar/kits-educativos/. [Accessed: 15-November-2020].
\bibitem{b12} Loomatix Ltd. 'Color Grab', 2020. [Online]. Available: http://www.loomatix.com/\#colorgrab. [Accessed: 15-November-2020].
\end{thebibliography}
\end{document}